\documentclass[conference,a4paper]{IEEEtran}

\usepackage{tikz}
\usepackage{amsmath}
\usepackage[english]{babel}

\usepackage{booktabs}
\usepackage{xurl}

\usepackage{tabularx}
\usepackage{xcolor}
\usepackage{siunitx}
\usepackage{setspace}
\usepackage{acronym}
\usepackage{rotating}
\usepackage{subcaption}
\usepackage{multirow}
\usepackage{listings}
\usepackage{pgfplots}
\usepackage{pgfplotstable}
\usepackage[pdfborder={0 0 0}]{hyperref}
\usepgfplotslibrary{external}
\usepackage{graphicx}
\usepackage{wrapfig}
\usepackage{etoolbox}

\usepackage[shortlabels]{enumitem} 

\usepackage{pgf-pie}
\newacro{VM}{virtual machine}
\newacro{AI}{artificial intelligence}
\newacro{FL}{federated learning}
\newacro{ML}{machine learning}
\newacro{DNN}{deep neural networks}
\newacro{MPU}{memory protection unit}
\newacro{FaaS}{Function as a Service}
\newacro{UUID}{Universally Unique Identifier}

\newcommand{\mg}[1]{\textcolor{purple}{#1}}

\newcommand{\kz}[1]{\textcolor{orange}{#1}}

\newcommand\bp[1]{\begin{itemize}\item \textcolor{blue}{#1} \end{itemize}}
\newcommand\bpi[1]{\begin{itemize}\item[]\bp{#1}\end{itemize}}
\newcommand{\code}[1]{\lstinline{#1}}
\usepackage{mathtools}
\usepackage{tikz-cd}
\usepackage{graphicx}
\usepackage{xspace}
\usepackage{footnote}
\pgfplotsset{compat=1.17}

\makeatletter
\patchcmd{\minted@colorbg}{\medskip}{}{}{}
\patchcmd{\endminted@colorbg}{\medskip}{}{}{}
\makeatother

\begin{document}
%-------------------------------------------------------------------------------
\bstctlcite{IEEEexample:BSTcontrol}

%don't want date printed
\date{}

%\title{\tinyfl: Remote Management of Federated Learning \ \\ on Resource-Constrained Edge IoT Devices}

\title{Model CBOR Serialization for Federated Learning
}

\author{\IEEEauthorblockN{Koen Zandberg}
\IEEEauthorblockA{Freie Universität Berlin\\
koen.zandberg@fu-berlin.de}
\and
\IEEEauthorblockN{Mayank Gulati}
\IEEEauthorblockA{Freie Universität Berlin\\
mayank.gulati@fu-berlin.de}
\and
\IEEEauthorblockN{Gerhard Wunder}
\IEEEauthorblockA{Freie Universität Berlin\\
g.wunder@fu-berlin.de}
\and
\IEEEauthorblockN{Emmanuel Baccelli}
\IEEEauthorblockA{Inria\\
emmanuel.baccelli@inria.fr}
}

%\author{Blind Review}

%-------------------------------------------------------------------------------

\maketitle

\begin{abstract}
%-Context
The typical federated learning workflow requires communication between a central server and a large set of clients synchronizing model parameters between each other.
% - problem
The current frameworks use communication protocols not suitable for resource-constrained devices and are either hard to deploy or require high-throughput links not available on these devices.
% - Solution
In this paper, we present %TinyFL, 
a generic message framework using CBOR for communication with existing federated learning frameworks optimised for use with resource-constrained devices and low power and lossy network links.
% - evaluation
We evaluate the resulting message sizes against JSON serialized messages where compare both with model parameters resulting in optimal and worst case serialization length, and with a real-world LeNet-5 model.
%We evaluate the resulting message sizes against similar JSON messages with both synthetic models and a real-world model.
% - measurements
Our benchmarks show that %the TinyFL
with our approach, messages are up to \qty{75}{\percent} smaller in size when compared to the JSON alternative.
% - Conclusion
% TODO
\end{abstract}
\section{Introduction}

%\bp{AI on edge devices is a nascent, very lively field}
%\bp{In this context, federated/distributed learning on low-power devices attractive because of reasons including privacy and customization...}
\Acf{AI}, and in particular \acf{ML} using \acf{DNN} have seen a spectacular development over the last decade.
In this context, nearly all verticals are substantially impacted by \ac{ML}, typically based on a data pipeline requiring the use of a model, i.e. a layered structure of algorithms which interpret data and make decisions based on that data. 
This model must first be trained in the learning phase, before it can be used for inference and put in production.

Recently, the TinyML community has been demonstrating the feasibility of executing model inference even on small microcontrollers, after these models have been trained and compressed on more powerful machines, as the learning phase typically requires enormous amounts of data and computing capacity.
Even more recently, however, the TinyML community explores the potential of learning on low-power microcontrollers~\cite{saha2022machine}.
In this context, a trend is \ac{FL} a machine learning paradigm where a model is trained across multiple decentralized devices without directly sharing their data with a central server.
\Ac{FL} and on-device training is attractive because of reasons including privacy, as raw privacy sensitive data does not leave the edge device, and customization with fine-tuning happening on-premises.

\subsection{Federated Learning Communication}
%\Ac{FL} is a machine learning paradigm where a model is trained across multiple decentralized devices without directly sharing their data with a central server. 
The typical workflow of \ac{FL} involves: (1) selecting participating devices; (2) sending a global model to each device; (3) training the model on local data; (4) sending local updates to a central server; (5) aggregating the updates to obtain a new global model; and (6) repeating the process until convergence.
The resulting model is then sent back to the participating devices for inference.
Throughout the \ac{FL} process, communication is crucial for managing the exchange of model updates between the central server and the clients.
It enables the synchronization of models, preserves privacy, and facilitates collaborative learning across distributed devices. Efficient and secure communication protocols are essential for successful FL implementation.

Typical \ac{FL} frameworks such as TensorFlow Federated~\cite{tensorflow-federated}, Flower~\cite{beutel2022flower}, etc. commonly rely on JSON~\cite{json} or gRPC~\cite{grpc} as part of their communication stack.
gRPC utilizes Protocol Buffers (Protobuf)~\cite{protobuf} as its message structure for data serialization and transmission.
Protobuf is a language-agnostic format that allows to define message types, fields, and optional values in a concise and efficient manner for serializing, deserializing, and manipulating the message objects.
However, setting up gRPC dependencies on severely resource-constrained devices can be challenging and human-readable JSON encoding is not optimized for embedded machine-to-machine communication.
To address this, alternative communication protocols are often used.
For instance, WebSockets~\cite{fette2011websocket} present a lightweight and real-time communication solution that is well-suited for such devices. 
MQTT~\cite{conf/comsware/HunkelerTS08}, specifically designed for resource-constrained environments, minimizes network bandwidth and power consumption while ensuring reliable \ac{FL} communication.
Furthermore, the widely supported HTTP/HTTPS protocols offer a standardized and user-friendly option for transmitting model updates in \ac{FL} workflows.

However with severely constrained devices these protocols are still too resource-intensive and not available.
Edge devices such as microcontrollers used in smart appliances have severely limited processing capabilities and memory, between \qtyrange{10}{100}{\mega\hertz} and \qtyrange{16}{256}{\kibi\byte}.
Furthermore the network link used by these devices is not optimized for the low latency and high throughput required by most protocols.
The typical stack of combining TCP and HTTPS with JSON or gRPC payloads puts a too heavy burden on these devices.
While MQTT is already used with \ac{FL}, other lightweight alternatives for these protocols used in the constrained device space such as MQTT-SN~\cite{mqtt-sn}, CoAP~\cite{rfc7252} and CBOR~\cite{rfc8949} are readily available, but have not been applied to the \ac{FL} space yet.
% \bp{typically on non-constrained machines/networks: use of BLABLABLA => Mayank?}
%\bp{on constrained devices/networks these are not applicable: need alternatives. => Koen?}

\subsection{Contributions}
In this paper, the work we present mainly consists in the following:
\begin{itemize}
\item We propose TinyFL a generic message data framework for \ac{FL}, on resource-constrained network nodes such as microcontroller-based devices. This framework enables efficient dissemination, monitoring and retrieval of \ac{ML} models among these nodes.
\item In order to minimize its footprint, TinyFL leverages the network protocol stacks and libraries typically present in the firmware of resource-constrained devices.
\item We evaluate the sizes of the messages incurred with TinyFL with a set of different \ac{ML} model sizes~\footnote{Source code for generating the results available at: \url{https://github.com/TinyPART/tinyfl-pemwn2023}}. We evaluate based both on idealized \ac{ML} models and on LeNet-5, a real-world model usable on microcontrollers;
\item We show that the CBOR-encoded messages used by TinyFL reduce the serialized size by up to \qty{75}{\percent} compared to a vanilla approach using JSON for instance. We show that the largest messages in the framework are only sent occasionally causing a minimal burden on the network link during the learning phase.
\end{itemize}

\if 0
\kz{KZ}
\bp{Design of an architecture for communication around federated learning between IoT edge devices
  - Focus on the message flow between the edge device and the aggregator
  - Leverage existing management solutions typically present on low-power ioT devices}
\bp{Implementation of Over-the-air model management framework, integrated in a popular low-power operating system (RIOT)}
\bp{Preliminary evaluation using FL for model X. Applicability: generality of the architecture goes beyond RIOT and beyond model X}
\fi
\section{Related Work}
Previous research, as demonstrated by existing work such as~\cite{10.1145/3462203.3475896} and TinyFedTL~\cite{kopparapu2021tinyfedtl}, has attempted to enable \ac{FL} training on small devices.
However, these studies lack a thorough analysis of computation and communication related to different model sizes, which is essential for understanding the practicality and suitability of \ac{FL} on tiny devices.
In contrast, EgdeML~\cite{PINYOANUNTAPONG2022109396} emphasizes the significance of a robust communication stack in \ac{FL} systems and points out the predominant focus on improving computation rather than optimizing the communication layer in existing research.
To address this challenge, we propose a modular framework that facilitates the interaction between the communication and computation layers, specifically designed for conducting \ac{FL} on constrained devices.

% Work on selecting a promising subset of clients among all available clients is done in~\cite{9249424}.

\section{Background}
\Ac{FL} is a research area that has gathered significant attention in recent years. It is a decentralized \ac{ML} paradigm where multiple devices or clients collaboratively train a global model without sharing their raw data. Instead, model updates are exchanged between the clients and a central server, allowing the server to aggregate and refine the shared model.

The concept of \ac{FL} was introduced in the seminal paper~\cite{mcmahan2017communication}. This paper laid the foundation for \ac{FL} by proposing a framework to train deep neural networks using decentralized data. The authors highlighted the challenges of \ac{FL}, such as communication efficiency and privacy preservation.

To address these challenges, subsequent research focused on improving communication efficiency in \ac{FL}. The paper by~\cite{konevcny2016federated} proposed strategies such as subsampling and quantization to reduce the amount of model updates exchanged between the clients and the server. 
% This approach helped alleviate communication bottlenecks and made \ac{FL} more scalable.

As \ac{FL} gained traction, system design considerations became crucial. Numerous advancements have been made extending the initial FL framework and addressing various aspects of deploying FL at scale. The comprehensive study ~\cite{kairouz2021advances} touchs upon issues related to scalability, fault tolerance, and security in large-scale \ac{FL} systems.

Our research focuses on using small devices as \ac{FL} clients to perform computational and communication tasks that contribute to the overall learning process. By utilizing tiny devices, our goal is to distribute workloads and improve the efficiency of \ac{ML}. We aim to tackle resource limitations and optimize communication protocols, enabling effective learning on these devices. Our ultimate aim is to contribute to the development of efficient and privacy-preserving distributed \ac{ML} systems.
% This approach allows for decentralized learning while preserving data privacy, as the devices collaborate and exchange model updates without sharing raw data.

% One of the widely used aggregation algorithms in \ac{FL} is Federated Averaging (FedAvg), introduced by~\cite{mcmahan2017communication}. FedAvg enables model aggregation by averaging the updates from multiple clients while minimizing communication overhead. This algorithm has become a key component in many \ac{FL} frameworks.

\subsection{Hardware and Basic Communication aspects}
The smallest of edge devices typically run on small constrained microcontrollers with low power interconnected via low throughput network links.
These devices have limited memory and computational processing power available, barely sufficient for small \ac{ML} models.
The resources available on these devices varies between \qtyrange{10}{100}{\mega\hertz} and \qtyrange{16}{256}{\kibi\byte},
with typical network links available such as IEEE802.15.4~\cite{9144691} limited to \qty{250}{\kilo\bit} with \qty{127}{\byte} maximum frame size.
In the usual network topology, these devices  connect via a gateway to a centralized and unconstrained entity acting as server.
Depending on the exact network topology, multiple network hops via intermediate (constrained) devices are necessary for a device to reach the gateway and contact the server.
These restrictions translate into a set of protocols and message formats optimized to deal with these constraints.

\subsubsection{CoAP}
CoAP~\cite{rfc7252}, as constrained counterpart to HTTP, provides a mechanism to interact with the clients in a constrained environment without putting significant burden on the wireless link and the client's processing power.
The protocol lends itself for RESTful interaction between clients and server over lossy and low power networks.
The observe mechanism allows for a CoAP client to subscribe to changes on a resource on a CoAP server, with publish/subscribe semantics, letting the CoAP server provide it with updates when the resource changes.

\subsubsection{CBOR}
CBOR~\cite{rfc8949} is a serialisation format similar to JSON with the main difference that it is a binary format.
It is optimised to result in a small serialised format and suitable for constrained links.
As CBOR dynamically extends the number of bytes required to encode numerical values, the encoded size heavily depends on the exact values encoded in the structure.
For example, low value integers up to \num{23} can be encoded in a single byte, with the number of bytes used increasing gradually to increase the required numerical range by the encoded value.
Additionally, CBOR also supports tagged data, a set of standardized tags are used to provide additional information on the following item.
This is used for example when encoding \acp{UUID}, which are encoded as a byte string with tag \num{37}.
Existing protocols such as SUIT~\cite{rfc9019} and SenML~\cite{rfc8428} make use of CBOR for serialising their messages when used with constrained devices.
\subsubsection{CDDL}
Describing the different CBOR data structures is possible using CDDL~\cite{rfc8610}.
The full expressiveness of CBOR, which exceeds that of JSON, can be unambiguously defined via CDDL.
It is both machine and human readable and can be used to validate CBOR data instances.
% maybe extend with minimal grammar define
\section{Scenario}

\begin{figure}
  \centering
  \includegraphics[width=.9\columnwidth]{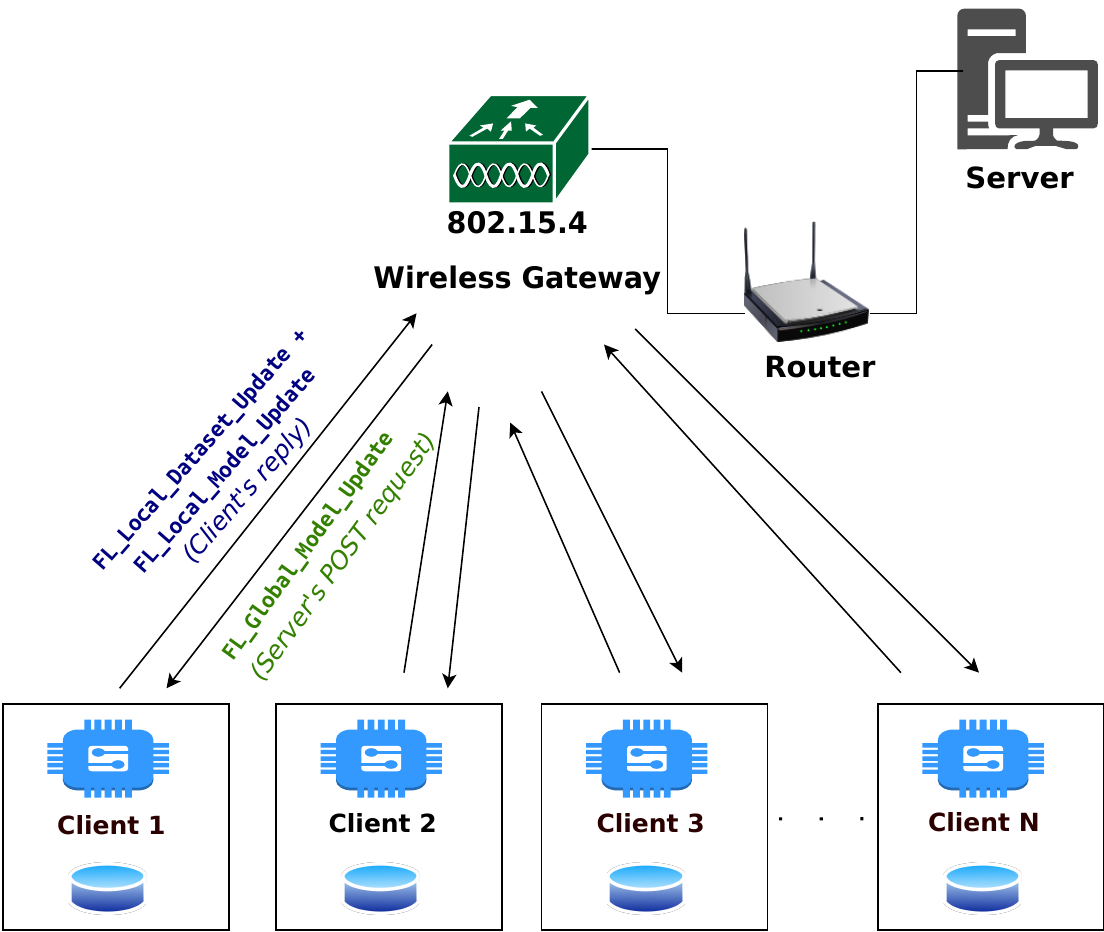}
  \caption{CoAP communication between \ac{FL} Server and Clients.}
  \label{fig:commflow}
  \vspace{-2em}
\end{figure}

In the scenario described here we assume that the \ac{FL} clients consists of a large number of microcontrollers, networked together over an already secured IEEE802.15.4 network using 6LoWPAN such as described by~\autoref{fig:commflow}.
Existing protocols are already available to provide mesh network capabilities and connect devices together in a secure way using protocols such as 6TiSCH~\cite{rfc9030}.
A number of software components are already present on the client firmware, reusing these for the \ac{FL} workflow would offer multiple benefits such as reduced memory usage. 
The firmware running on the microcontrollers is assumed to include a number of existing modules for protocols among at least: CoAP to provide REST-like communication with a server and CBOR for serializing messages.
Furthermore, it is assumed that there is a protocol in place for system-level management of the clients, such as CORECONF~\cite{ietf-core-comi-12} to provide the initial on-boarding and configuration of clients.
While we limit the scenario to using CoAP, other protocols optimised for constrained environments such as MQTT-SN could also be used in the workflow without changing the message formats.

\if{0}
\bp{Constrained network link between the edge devices}
\bp{Typically: Secure communication handled by existing protocols, (includes registration of devices)}
\bpi{LAKE}
\bpi{OSCORE}
\bpi{COSE}
\bp{Typically: Low processing capability on the edge devices}
\bp{Aggregator is unconstrained}

\subsection{machine learning aspects}
\mg{MG}
\bp{what needs to be updated?} 
\bp{fetching parameters from server for model initialization and writing local updates}

\bp{how often?} 
\bpi{time period based or event based}
\bp{which traffic pattern?}
\bpi{star topology : clients + aggregator}
  
\fi

\section{Architecture}
The full \ac{FL} architecture is a multi-step process started and orchestrated by a central server.
As the first step, the server sets up the orchestration process, defining the number of participating clients for training, minimum fraction of these clients required for aggragation, the number of \ac{FL} rounds to be performed, and the stop condition for individual clients. Furthermore, the server decides a minimum of data samples required in order to accomplish local training. This is necessary to ensure that the local model is trained suffiently enough to make its contribution count for global model aggregation.
% Furthermore, the server decides the number of epochs for local training that should be executed at each client, based on either a fixed-size local dataset or a minimum number of data samples in the case of streaming data.

Once the orchestration is configured, the server initializes a global model. At the beginning of each round, a global model is sent to each participating client device. Each client then independently trains the model by performing several iterations of gradient update steps using their local dataset. The clients have two types of datasets: a training set and a separate validation set, used for training and evaluating models, respectively. Meanwhile, the server observes if a particular client has been trained sufficiently. In response, clients reply with the number of data samples seen so far during training and other performance metrics, such as loss and accuracy, which can be later used to determine the stopping condition for an individual client when the metrics show no significant improvement. During our experimental process, we incorporated a stopping condition based on the comparison between the validation loss and the training loss. Specifically, when the validation loss became lower than the training loss, the server would intervene and halt the training process for that specific client.

After a sufficient number of clients respond, the server will request the trained model parameters from these clients, along with the sizes of their respective datasets. This is done in order to perform model aggregation, such as weighted model averaging, as seen in the FedAvg~\cite{mcmahan2017communication} approach.

\subsection{Communication}

\begin{figure}
  \centering
  \includegraphics[scale=0.7]{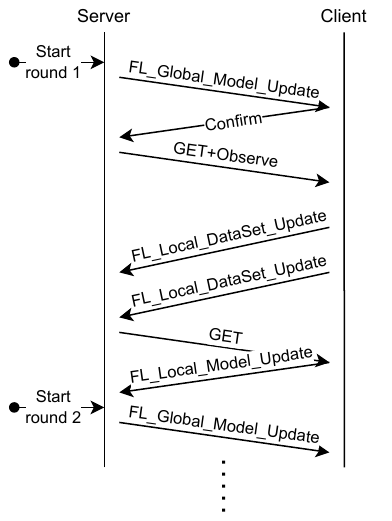}
  \caption{CoAP communication timing diagram between \ac{FL} Server and a single client.}
  \label{fig:networktiming}
\end{figure}

As described above, the \ac{FL} workflow is a multi-step process over multiple rounds.
This requires multiple requests and responses from the server to the clients.
An overview of the communication flow between the server and a single client is shown in \autoref{fig:networktiming}.

\subsubsection{Client configuration}

The first requests from the server to the clients is a POST request with the initial or new global model.
This request is CBOR-encoded and is formatted as shown in Listing 1. %~\ref{lst:global}.
% TODO: validate CDDL
%\begin{listing}[!ht]
%\inputminted[lastline=13]{cddl}{cddl/full.txt}
\begin{lstlisting}[caption=\textit{CDDL description of a global model update payload}]
FL_Global_Model_Update = [
  fl-model-identifier,
  fl-model-round,
  fl-model-params,
  fl-continue-training : bool
]

fl-model-identifier = #6.37(bstr)
fl-model-round = uint
fl-model-params /= [+ float]
fl-model-params /= ta-float16le
fl-model-params /= ta-float32le
fl-model-params /= ta-float64le
\end{lstlisting}
%\caption{CDDL description of a global model update payload}
\label{lst:global}
%\end{lstlisting}

The data in the request contains an \ac{UUID} (the \code{fl-model-identifier}, with the \ac{UUID} encoded as tagged byte string) to identify the model and to allow for multiple models simultaneously on the clients.
A round number is included to version the global model between rounds. 
Multiple encodings are allowed for serialising the list of parameters, the simplest being an array of floating point numbers, the CBOR encoding allowing for dynamically scaling between half-float, float and doubles depending on the required accuracy for the provided values.
The four other encoding options can be used when the number of parameters all use the same type and can be encoded as an homogeneous type following the CBOR typed array format~\cite{rfc8742}.
In this format the parameters are encoded as a byte string with each parameter concatenated in the representation specified by a CBOR tag around the byte string.
Finally the request includes a flag to specify whether the client should start training a new local model based on the supplied global model or run in inference-only mode.

%\begin{listing}[!ht]
%\inputminted[firstline=15, lastline=23]{cddl}{cddl/full.txt}
\begin{lstlisting}[caption=\textit{CDDL description of a client update payload.}]
FL_Local_DataSet_Update = [
  fl-local-dataset-size : uint,
  ? fl-model-metadata,
]

fl-model-metadata = (
  fl-local-model-train-loss: float
  fl-local-model-val-loss : float
)
\end{lstlisting}
\label{lst:client-update}
%\caption{CDDL description of a client update payload}
%  \vspace{-1em}
%\end{listing}

After starting the new training period on the clients, the server submits the CoAP GET with observe request to the clients.
The level of training required by the server is submitted as query parameter in the CoAP GET request.
Via the observe mechanism, this request provides the server with a notification as soon as a client reached the the minimum level of training required by the client. % Please check phrasing here
The format in CDDL used by the clients for the replies are shown in Listing 2. %~\ref{lst:client-update}.

After a sufficient number of clients have all gathered sufficient local data for training the local model, the server queries the selected clients for their local model update.
This is done via a GET request and the clients reply with a structure shown in Listing 3. %~\ref{lst:client-model}.

The update contains the same \ac{UUID} and round number used with the initial global model update described in Listing 1. %~\ref{lst:global}.
As main data, the client provides the local dataset size used to train the model with, and the local model itself as list of floating point numbers.

\if{0}
\kz{KZ}
\bp{Describe communication flow between nodes and central aggregator}
\bp{CoAP with CBOR for encoding and transport}
\bp{Describe messages:}
\bpi{Param updates from nodes to aggregator}
\bpi{Model updates from aggregator to nodes}

\bp{Use LAKE etc for security at network layer and registration.}
\fi

%\begin{listing}[!ht]
%\inputminted[firstline=25]{cddl}{cddl/full.txt}
\begin{lstlisting}[caption=\textit{CDDL description of a client model payload.}]
FL_Local_Model_Update = [
  fl-model-identifier,
  fl-model-round,
  fl-model-params,
  fl-model-metadata,
]
\end{lstlisting}
%\caption{CDDL description of a client model payload}
\label{lst:client-model}
%  \vspace{-1em}
%\end{listing}
\section{Evaluation}
\subsection{Measurement setup}
To evaluate the message framework proposed here, we evaluate the different messages on a two metrics.
\subsubsection{Message size}
The main metric for evaluation is the message size in bytes. 
As the scenario involves severely constrained network links, keeping the message size below the maximum frame size of the link (\qty{127}{\byte}) prevents requiring multiple radio transmissions for a single message.
As resulting size of the CBOR encoding depends a lot on the value of the data, we provide both optimistic and pessimistic values for the message sizes.
The measurements were done using three different simulated model sizes: a small model of 4 floating point numbers, an intermediate model size of 1000 floating point numbers and a large model with 10 000 floating point numbers.
We also measure the size of the messages when they are encoded as Protobuf and as minified JSON messages, to show the reduction in message size due to the CBOR encoding.
For the floating point numbers, the value \code{1.0} was chosen, as this requires the least amount of encoding size with JSON.
The measured size for JSON represents the minimal size that can be achieved and should be compared against the best case with the CBOR encoding.

\subsubsection{Message interval}
The rate at which messages are send between the server and clients can be classified.
Preferably larger messages are only required in exceptional cases, in contrast to the updates from clients to server, which should be small as they happen often.
The exact interval of the messages depend on the training configuration.

\subsection{Measurements}

%%%%%%%%%%%%%%%%%%%%%%%%%%%%%

\begin{table*}
  \centering
  \begin{tabular}{lrrrrr}
    \toprule
  Message & Model Size & CBOR Best & CBOR Worst & Protobuf & JSON\\
    \midrule
    \code{FL_Local_DataSet_Update} &  & \SI{8}{\byte} & \SI{28}{\byte} & \qty{22}{\byte} & \SI{11}{\byte} \\
    \midrule
    \multirow{3}{13em}{\code{FL_Global_Model_Update}} & \num{4} & \SI{33}{\byte} & \SI{67}{\byte} & \SI{40}{\byte} & \SI{65}{\byte} \\
                                                      & \num{1000} & \SI{2027}{\byte} & \SI{9033}{\byte} & \SI{4025}{\byte} & \SI{4049}{\byte} \\
                                                      & \num{10000} & \SI{20025}{\byte} & \SI{90033}{\byte} & \qty{40026}{\byte} & \SI{40049}{\byte} \\
    \midrule
    \multirow{3}{13em}{\code{FL_Local_Model_Update}}  & \num{4} & \SI{38}{\byte} & \SI{84}{\byte} & \SI{58}{\byte} & \SI{68}{\byte} \\
                                                      & \num{1000} & \SI{2032}{\byte} & \SI{9050}{\byte} & \SI{4043}{\byte} & \SI{4052}{\byte} \\
                                                      & \num{10000} & \SI{20032}{\byte} & \SI{90050}{\byte} & \qty{40044}{\byte} & \SI{40052}{\byte} \\
    \bottomrule
  \end{tabular}
\caption{Message sizes when encoded as CBOR, Protobuf and JSON, for varying model sizes}
\label{tbl:msgsizes}
\end{table*}

%%%%%%%%%%%%%%%%%%%%%%%%%%%%%

\begin{table*}
  \centering
  \begin{tabular}{lrrr}
    \toprule
  Message & CBOR & ProtoBuf & JSON\\
    \midrule
    \code{FL_Global_Model_Update} & \SI{177733}{\byte} & \qty{177730}{\byte} & \qty{928171}{\byte} \\
    \code{FL_Local_Model_Update} & \SI{177738}{\byte} & \qty{177748}{\byte} & \SI{928168}{\byte} \\
    \bottomrule
  \end{tabular}
\caption{Message sizes when encoding the LeNet-5 model as CBOR and JSON}
\label{tbl:msgsizelenet}
  \vspace{-1.5em}
\end{table*}

%%%%%%%%%%%%%%%%%%%%%%%%%%%%%

\subsubsection{Message size}
The messages sizes for different model sizes is shown in \autoref{tbl:msgsizes}.
First, the \code{FL_Local_DataSet_Update} message that contains the updates from the client is between \SI{8}{\byte} and \SI{28}{\byte} with CBOR, depending on the exact values used.
While the JSON encoded variant is slighly larger than the best case CBOR encoding, both are suitably small to fit inside a single frame on a constrained network.
The protobuf message is in same range as the CBOR message size.
For CBOR, the message sizes for messages that contain a small model are between \SI{33}{\byte} and \SI{84}{\byte}.
With larger model sizes between \num{1000} and \num{10000} parameters, the message sizes increase with the same magnitude.
The best case for CBOR is using half floats, which is visible when comparing with the Protobuf sizes, as Protobuf always uses at least \qty{32}{\bit} floats.
As a JSON-encoded float inside an array needs 4 characters and a CBOR-encoded half float inside a tagged array needs 2 bytes, the size of the CBOR-encoded messages approximates \SI{50}{\percent} of the JSON-encoded messages.
These messages are no longer small enough to fit inside a single transmission frame and require the CoAP blockwise transfer mechanism to get fully transferred.
The \code{FL_Local_DataSet_Update} size is independent of the number of the model parameters, it does not contain the model parameters.

\subsubsection{Message interval}
% Maybe restructure this to a different section
Looking at how often the different messages need to be transferred we can distinguish two update frequencies.
The \code{FL_Local_DataSet_Update} will be transmitted relative often and is unique per client during the training round.
To save throughput and in turn power on the clients, it is important that this message is as small as possible.
At \SI{28}{\byte} maximum it will always fit inside a single transmission from the client.

The \code{FL_Global_Model_Update} depends on the size of the model and can potentially be a large message that needs to be distributed to all clients.
However it only needs to be distributed once per round.
As all clients need the same message, strategies to disseminate this message through the network can be applied, such as using a single multi-cast message reaching all clients.
The \code{FL_Local_Model_Update} from the clients to the server is also transferred only once per round.
Each selected client will have to transfer the \code{FL_Local_Model_Update} message to the server with their locally trained model parameters.
As this message is unique per client and contains the parameters of the model, it puts the largest burden on the network.
However, not all clients have to transfer this message, but only the clients selected by the server.

\subsubsection{Real world example}
Finally we compare the message sizes when using the LeNet-5~\cite{lecun1998gradient} model (approx. \num{45000} parameters) in the message structures.
The results are shown in \autoref{tbl:msgsizelenet}.
The \code{FL_Local_DataSet_Update} message has been omitted as the size is independent of the model size and is identical to the previous shown measurements.
Visible is the significant gain by encoding the messages in CBOR.
Where previous tables compared a best case CBOR messages with a best case JSON messages in terms of size, here the average case with real-world values is shown.
Compared to JSON, the message size is around \qty{24}{\percent} of the size.

\section{Discussion and Future Work}
With the minimal encoding structure from CBOR, flexibility is available to encode the models in their smallest size possible.
The CBOR tagged-array format results in a message size at most the same size as the as the required memory for the model parameters on the device.
However depending on the specific model parameter values it can result in a size smaller than the on-device representation.

In future work, it could be valuable to explore the possibility of personalizing the global model on the client side before deploying it for inference. Allowing clients to fine-tune the global model using their local data and domain-specific knowledge could lead to improved performance and tailored predictions. Additionally, it would be worth investigating the capability of discarding server updates and relying solely on local updates if the performance of the global model deteriorates. This approach would require monitoring the model's performance at the client level and selectively using local updates to maintain or improve performance. However, integrating these capabilities into constrained \ac{FL} system requires careful consideration of factors such as compute resource availability, model complexity, and communication overhead. Developing novel algorithms and protocols will be essential to enable efficient and effective collaboration between clients and the server. Moreover, our framework could be used to transfer of partial models, allowing for flexible and efficient transfer learning scenarios, where only specific layers or components need to be exchanged.

% The message format could be adjusted to contain not only the full model, but also allow for transferring partial models.
% This can significantly reduce the size of the messages and reduce the burden on the network links.
% One approach for this is to modify the \code{fl-model-params} structure into a map of arrays, the key of the map specifying the part of the model to which the parameters apply.
% with this multiple layers, but not all layers, of a model can be updated in a transfer.
\section{Conclusion}

Our message and communication framework provides a highly effective approach for implementing \ac{FL} on microcontrollers, surpassing the limitations of traditional Protobuf and JSON-based methods and communication protocols. There are several key advantages to our framework. Firstly, it strikes an optimal balance between message size and frequency, ensuring efficient and resource-conscious communication. Secondly, it seamlessly integrates with existing IoT management systems, leveraging widely adopted network stack building blocks. This enables easy integration and scalability within larger IoT infrastructures. This capability enhances the versatility and adaptability of \ac{FL} on microcontrollers, unlocking new possibilities for distributed learning applications.
\section*{Acknowledgements}
The research leading to these results partly received funding from the
MESRI-BMBF German/French cybersecurity program under
grant agreements No. ANR-20-CYAL-0005 and 16KIS1395K.
The paper reflects only the authors’ views. MESRI and BMBF
are not responsible for any use that may be made of the
information it contains.

\bibliography{IEEEabrv,bibliography}

\begin{thebibliography}{10}
\providecommand{\url}[1]{#1}
\csname url@samestyle\endcsname
\providecommand{\newblock}{\relax}
\providecommand{\bibinfo}[2]{#2}
\providecommand{\BIBentrySTDinterwordspacing}{\spaceskip=0pt\relax}
\providecommand{\BIBentryALTinterwordstretchfactor}{4}
\providecommand{\BIBentryALTinterwordspacing}{\spaceskip=\fontdimen2\font plus
\BIBentryALTinterwordstretchfactor\fontdimen3\font minus
  \fontdimen4\font\relax}
\providecommand{\BIBforeignlanguage}[2]{{%
\expandafter\ifx\csname l@#1\endcsname\relax
\typeout{** WARNING: IEEEtran.bst: No hyphenation pattern has been}%
\typeout{** loaded for the language `#1'. Using the pattern for}%
\typeout{** the default language instead.}%
\else
\language=\csname l@#1\endcsname
\fi
#2}}
\providecommand{\BIBdecl}{\relax}
\BIBdecl

\bibitem{saha2022machine}
S.~S. Saha, S.~S. Sandha, and M.~Srivastava, ``Machine learning for
  microcontroller-class hardware-a review,'' \emph{IEEE Sensors Journal}, 2022.

\bibitem{tensorflow-federated}
\BIBentryALTinterwordspacing
``{TensorFlow Federated}.'' [Online]. Available:
  \url{https://www.tensorflow.org/federated}
\BIBentrySTDinterwordspacing

\bibitem{beutel2022flower}
D.~J. Beutel \emph{et~al.}, ``Flower: A friendly federated learning
  framework,'' 2022.

\bibitem{json}
E.~International, ``Ecma-404—the json data interchange format,'' 2013.

\bibitem{grpc}
\BIBentryALTinterwordspacing
``{gRPC}.'' [Online]. Available: \url{https://grpc.io/}
\BIBentrySTDinterwordspacing

\bibitem{protobuf}
\BIBentryALTinterwordspacing
``Protocol buffers.'' [Online]. Available: \url{https://protobuf.dev/}
\BIBentrySTDinterwordspacing

\bibitem{fette2011websocket}
I.~Fette and A.~Melnikov, ``The websocket protocol,'' RFC 6455, 2011.

\bibitem{conf/comsware/HunkelerTS08}
U.~Hunkeler \emph{et~al.}, ``Mqtt-s - a publish/subscribe protocol for wireless
  sensor networks.'' in \emph{COMSWARE}, S.~Choi, J.~Kurose, and
  K.~Ramamritham, Eds.\hskip 1em plus 0.5em minus 0.4em\relax IEEE, 2008, pp.
  791--798.

\bibitem{mqtt-sn}
A.~Stanford-Clark and H.~L. Truong, ``Mqtt for sensor networks (mqtt-sn)
  protocol specification,'' \emph{IBM Corporation version}, vol.~1, no.~2, pp.
  1--28, 2013.

\bibitem{rfc7252}
\BIBentryALTinterwordspacing
Z.~Shelby, K.~Hartke, and C.~Bormann, ``{The Constrained Application Protocol
  (CoAP)},'' RFC 7252, Jun. 2014. [Online]. Available:
  \url{https://www.rfc-editor.org/info/rfc7252}
\BIBentrySTDinterwordspacing

\bibitem{rfc8949}
\BIBentryALTinterwordspacing
C.~Bormann and P.~E. Hoffman, ``{Concise Binary Object Representation
  (CBOR)},'' RFC 8949, Dec. 2020. [Online]. Available:
  \url{https://www.rfc-editor.org/info/rfc8949}
\BIBentrySTDinterwordspacing

\bibitem{10.1145/3462203.3475896}
M.~M. Grau \emph{et~al.}, ``On-device training of machine learning models on
  microcontrollers with a look at federated learning,'' in \emph{Proceedings
  ACM GoodIT}, 2021, p. 198–203.

\bibitem{kopparapu2021tinyfedtl}
K.~Kopparapu and E.~Lin, ``Tinyfedtl: Federated transfer learning on tiny
  devices,'' \emph{arXiv preprint arXiv:2110.01107}, 2021.

\bibitem{PINYOANUNTAPONG2022109396}
P.~Pinyoanuntapong \emph{et~al.}, ``Edgeml: Towards network-accelerated
  federated learning over wireless edge,'' \emph{Computer Networks}, vol. 219,
  p. 109396, 2022.

\bibitem{mcmahan2017communication}
B.~McMahan \emph{et~al.}, ``Communication-efficient learning of deep networks
  from decentralized data,'' in \emph{Artificial intelligence and
  statistics}.\hskip 1em plus 0.5em minus 0.4em\relax PMLR, 2017, pp.
  1273--1282.

\bibitem{konevcny2016federated}
J.~Kone{\v{c}}n{\`y} \emph{et~al.}, ``Federated learning: Strategies for
  improving communication efficiency,'' \emph{arXiv preprint arXiv:1610.05492},
  2016.

\bibitem{kairouz2021advances}
P.~Kairouz \emph{et~al.}, ``Advances and open problems in federated learning,''
  \emph{Foundations and Trends{\textregistered} in Machine Learning}, vol.~14,
  no. 1--2, pp. 1--210, 2021.

\bibitem{9144691}
``Ieee standard for low-rate wireless networks,'' \emph{IEEE Std
  802.15.4-2020}, pp. 1--800, 2020.

\bibitem{rfc9019}
\BIBentryALTinterwordspacing
B.~Moran \emph{et~al.}, ``{A Firmware Update Architecture for Internet of
  Things},'' RFC 9019, Apr. 2021. [Online]. Available:
  \url{https://www.rfc-editor.org/info/rfc9019}
\BIBentrySTDinterwordspacing

\bibitem{rfc8428}
\BIBentryALTinterwordspacing
C.~F. Jennings \emph{et~al.}, ``{Sensor Measurement Lists (SenML)},'' RFC 8428,
  Aug. 2018. [Online]. Available: \url{https://www.rfc-editor.org/info/rfc8428}
\BIBentrySTDinterwordspacing

\bibitem{rfc8610}
\BIBentryALTinterwordspacing
H.~Birkholz, C.~Vigano, and C.~Bormann, ``{Concise Data Definition Language
  (CDDL): A Notational Convention to Express Concise Binary Object
  Representation (CBOR) and JSON Data Structures},'' RFC 8610, Jun. 2019.
  [Online]. Available: \url{https://www.rfc-editor.org/info/rfc8610}
\BIBentrySTDinterwordspacing

\bibitem{rfc9030}
\BIBentryALTinterwordspacing
P.~Thubert, ``{An Architecture for IPv6 over the Time-Slotted Channel Hopping
  Mode of IEEE 802.15.4 (6TiSCH)},'' RFC 9030, May 2021. [Online]. Available:
  \url{https://www.rfc-editor.org/info/rfc9030}
\BIBentrySTDinterwordspacing

\bibitem{ietf-core-comi-12}
\BIBentryALTinterwordspacing
M.~Veillette \emph{et~al.}, ``{CoAP Management Interface (CORECONF)},''
  Internet Engineering Task Force, Internet-Draft draft-ietf-core-comi-12, Mar.
  2023. [Online]. Available:
  \url{https://datatracker.ietf.org/doc/draft-ietf-core-comi/12/}
\BIBentrySTDinterwordspacing

\bibitem{rfc8742}
\BIBentryALTinterwordspacing
C.~Bormann, ``{Concise Binary Object Representation (CBOR) Sequences},'' RFC
  8742, Feb. 2020. [Online]. Available:
  \url{https://www.rfc-editor.org/info/rfc8742}
\BIBentrySTDinterwordspacing

\bibitem{lecun1998gradient}
Y.~LeCun, L.~Bottou, Y.~Bengio, and P.~Haffner, ``Gradient-based learning
  applied to document recognition,'' \emph{Proceedings of the IEEE}, vol.~86,
  no.~11, pp. 2278--2324, 1998.

\end{thebibliography}
%%%%%%%%%%%%%%%%%%%%%%%%%%%%%%%%%%%%%%%%%%%%%%%%%%%%%%%%%%%%%%%%%%%%%%%%%%%%%%%%
\end{document}